\documentclass{ws-procs9x6}

\usepackage{hyperref}	

\begin{document}

\title{ Schwinger-Dyson Equations in\\
Group Field Theories of Quantum Gravity
}

\address{"The XXIX International Colloquium 
on\\ Group-Theoretical Methods in 
Physics"\\
Chern Institute of Mathematics\\ 
Tianjin,  China\\
August 20-26, 2012}

\author{Thomas Krajewski}

\address{Centre de Physique Th\'eorique\\
Campus de Luminy\\
F-13288 Marseille cedex 09 France\\
E-mail: Thomas.Krajewski@cpt.univ-mrs.fr}

\begin{abstract}
In this talk, we elaborate on the operation of graph contraction introduced by Gurau in his study of the Schwinger-Dyson equations. After a brief review of colored tensor models, we identify the Lie algebra appearing in the Schwinger-Dyson equations as a Lie algebra associated to a Hopf algebra of the Connes-Kreimer type. Then, we show how this operation also leads to an analogue of the Wilsonian flow for the effective action. Finally, we sketch how this formalism may be adapted to group field theories.

\end{abstract}


\bodymatter

\section{A very short overview of colored tensor models}

Colored models have been introduced by Gurau in the context of group field theory\cite{colored}. These models admit  a perturbative expansion dominated by spherical topologies\cite{complete} and are at the roots of the first renormalizable group field theory\cite{renormalizability}. Here, we follow closely the formalism introduced by Gurau in the study of the Schwinger-Dyson equations\cite{GurauSD}. In dimension $D$, the basic  dynamical variables of a colored tensor model are two complex conjugate rank $D$ tensors $M_{i_{1}\dots i_{D}}$ and  $\overline{M}_{i_{1}\dots i_{D}}$, with $i_{k}\in\left\{1,\dots,N\right\}$.  To any bipartite graph with black and white vertices of valence $D$ and a proper coloring with $D$ colors, we associate a $\big({\mathrm U}(N)\big)^{D}$ invariant  denoted $\mathrm{Tr}_{\Gamma}(M,\overline{M})$ constructed as follows. To any white (resp. black) vertex we associate a tensor $M_{i_{1}\dotsi_{D}}$ (resp. $\overline{M}_{i_{1}\dots i_{D}}$) and contract the $n^{\mathrm{th}}$ indices of the tensors following the edges of color $n$ in $\Gamma$. For instance,
 \begin{equation}
\parbox{2cm}{ \includegraphics[width=2cm]{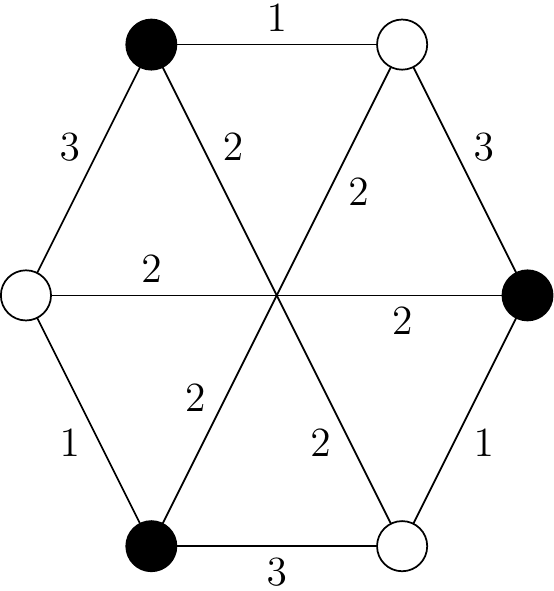}}
\rightarrow
\sum_{i_{a},\,j_{b},\,k_{c}}
\overline{M}_{i_{1}i_{2}i_{3}}\overline{M}_{j_{1}j_{2}j_{3}}\overline{M}_{k_{1}k_{2}k_{3}}
M_{i_{1}k_{2}j_{3}}M_{j_{1}i_{2}k_{3}}M_{k_{1}j_{2}i_{3}}
 \end{equation} 
The main object of interest is the generating function
\begin{equation}
{\cal W}[\lambda_{\Gamma}]=\log
\int[DMD\overline{M}]\,\exp\bigg\{\sum_{\Gamma}\frac{\lambda_{\Gamma}}{C_{\Gamma}}\mbox{Tr}_{\Gamma}(M,\overline{M})\bigg\},\label{generating}
\end{equation}
where $\int[DMD\overline{M}]$ is the Lebesgue measure on the real and imaginary part of the tensors and $C_{\Gamma}$ is the cardinal of the automorphism group of $\Gamma$. The integral ${\cal W}[\lambda_{\Gamma}]$ is defined  as a formal power series as follows. We single out the contribution of the dipole graph
\begin{equation}
\parbox{2cm}{\includegraphics[width=2cm]{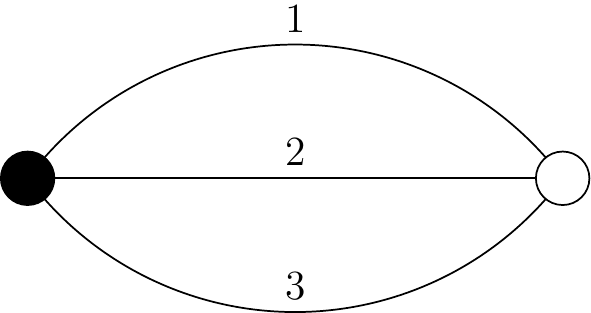}}
\rightarrow
\sum_{i_{a}}\overline{M}_{i_{1},i_{2},i_{3}}M_{i_{1}i_{2}i_{3}}
\end{equation}
whose exponential defines a Gau\ss ian measure on tensors. Then, we use Wick's theorem to expand  ${\cal W}[\lambda_{\Gamma}]$ as a formal power series in the coefficients $\lambda_{\Gamma}$, where the number of vertices $\Gamma$ is strictly greater than 2. From a geometric perspective, each graph $\Gamma$ represents a triangulation of a space of dimension $D$, with $D$-simplexes on its vertices glued together along their boundary $(D\!-\!1)$-simplexes following the edges of $\Gamma$. The Feynman graphs appearing in the expansion of ${\cal W}[\lambda_{\Gamma}]$ are associated with cellular decompositions of $(D\!+\!1)$-dimensional spaces.

\section{Graph contraction and Schwinger-Dyson equations }

In order to derive the Schwinger-Dyson equations for tensor models \cite{GurauSD}, it is convenient to define the contraction of a pair of vertices $v,\overline{v}$ of different colors in a non necessarily connected, bipartite colored graph $\Gamma$ of degree $D$. It is the graph $\Gamma/\overline{v}v$ obtained by removing the vertices $v$ and $\overline{v}$ and attaching the resulting free half-edges of identical colors. For example, in $D=3$,

\medskip

\hskip-0.8cm
\begin{tabular}{ccc}
\parbox{3cm}{\includegraphics[width=3cm]{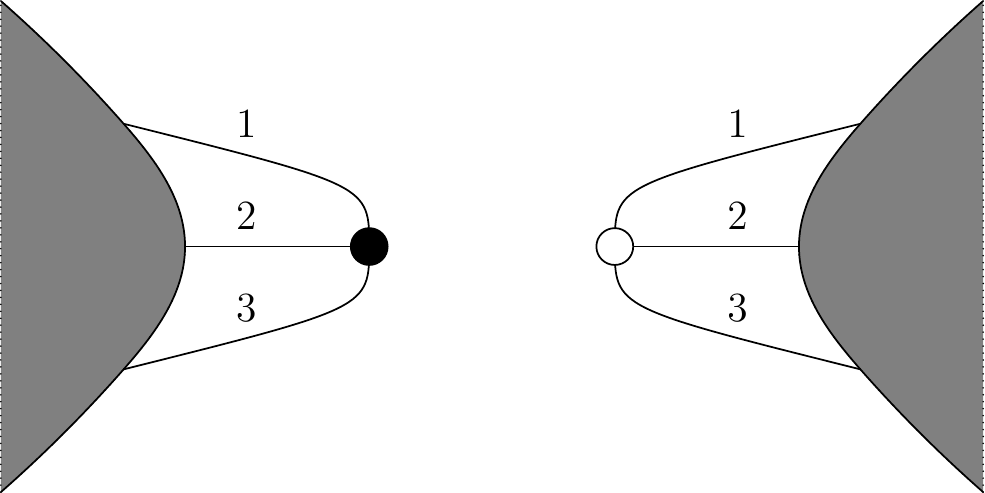}}$\rightarrow$
\parbox{1.8cm}{\includegraphics[width=1.8cm]{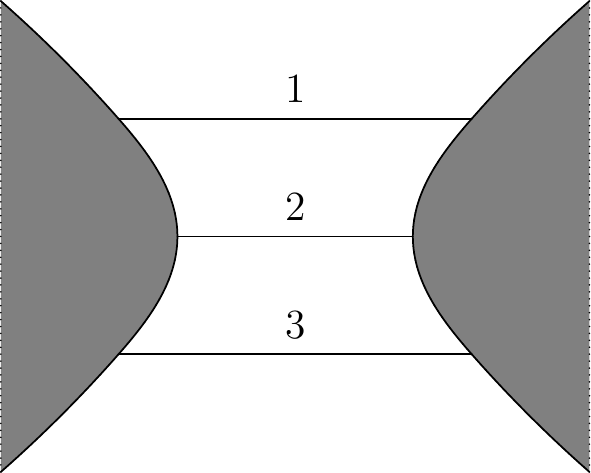}}&\quad&\parbox{3cm}{\includegraphics[width=3cm]{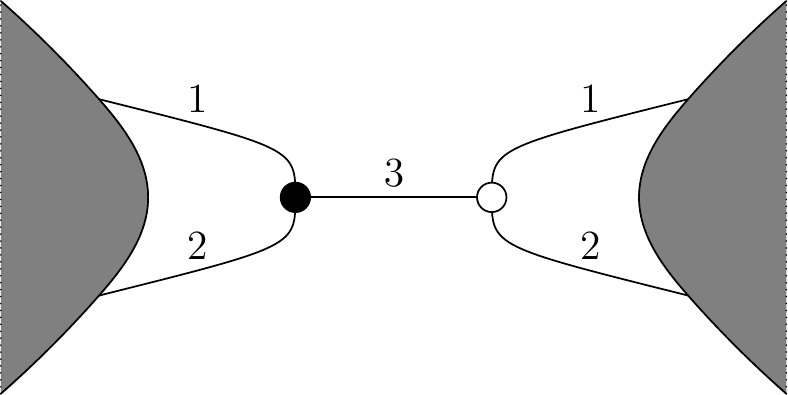}}$\rightarrow$
\parbox{1.8cm}{\includegraphics[width=1.8cm]{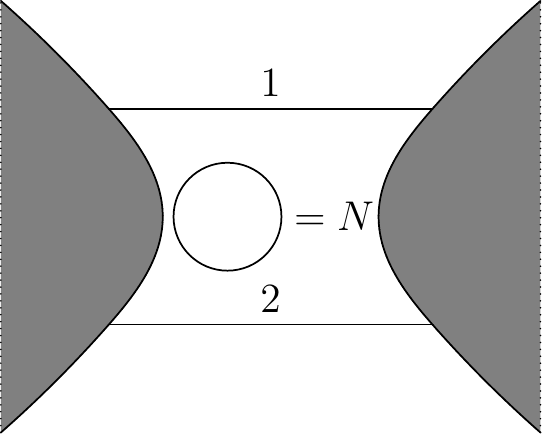}}\\
vertices not sharing an edge&&vertices sharing an edge
\end{tabular}

\medskip

In the second illustration, $v$ and $\overline{v}$ share an edge, which results in a extra loop in the contracted graph $\Gamma/\overline{v}v$. This extra loop simply yields a factor of $N$ and can be omitted, In all case, $\Gamma/\overline{v}v$ is also a colored bipartite graph of degree $D$. However, $\Gamma/\overline{v}v$ may not be connected even if $\Gamma$ is.

Then the Schwinger-Dyson equations for colored tensor models\cite{GurauSD} read
 {\begin{equation}
\bigg( \sum_{v=\circ\in\Gamma_{0}\,}
\prod_{k\,\mbox{\tiny connected}\atop\mbox{\tiny components}}
\frac{\partial}{\partial\lambda_{[\Gamma_{0}/\overline{v}_{0}v]^{}_{k}}}+\sum_{\Gamma\,\mbox{\tiny connected},\, v=\circ\in\Gamma} \lambda_{\Gamma}
 \frac{\partial}{\partial\lambda_{(\Gamma_{0}\Gamma)/\overline{v}_{0}v}}\bigg){\cal W}[\lambda_{\Gamma}]=0\label{SD}
 \end{equation}
with the the convention that $\frac{\partial}{\partial\lambda_{[\Gamma_{0}/\overline{v}_{0}v]^{}_{k}}}=N$ if $[\Gamma_{0}/\overline{v}_{0}v]^{}_{k}$ is a loop without vertex. To derive these equation we choose a graph $\Gamma$ and a black vertex $\overline{v}_{0}\in\Gamma_{0}$ and perform the infinitesimal change of variable $M\rightarrow M+\delta_{(\Gamma,\overline{v}_{0})}M$ in the integral defining ${\cal W}[\lambda_{\Gamma}]$, 
with $\delta_{(\Gamma,\overline{v}_{0})}M$ equal to $\mathrm{Tr}_{\Gamma}(M,\overline{M})$ with the contribution of vertex $\overline{v}_{0}$ removed. The first term in \eqref{SD} arises from the Jacobian whereas the second one comes from the variation of the integrand. It is worthwhile to notice that Schwinger-Dyson equations may used to determine the large $N$ limit\cite{Bonzom}.

Geometrically, Schwinger-Dyson equations can be represented as
\begin{equation}
\sum_{v\in \Gamma_{0}\atop v\neq v_{0}}\left\langle\parbox{3cm}{\includegraphics[width=3cm]{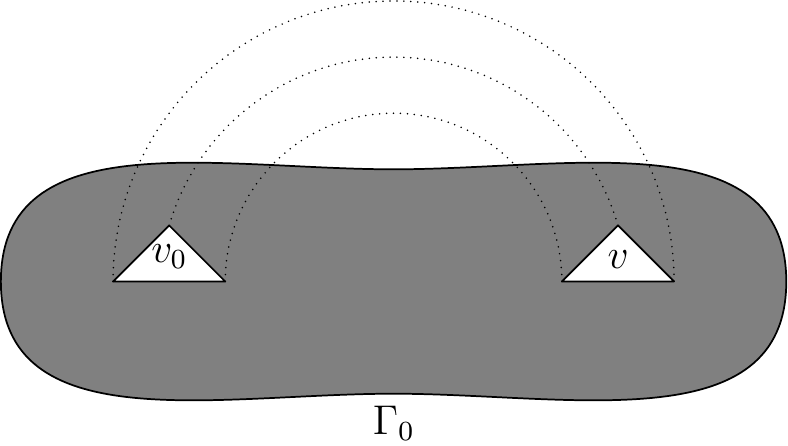}}\right\rangle\quad+\quad
\sum_{\Gamma\atop v\in\Gamma}\left\langle\parbox{3cm}{\includegraphics[width=3cm]{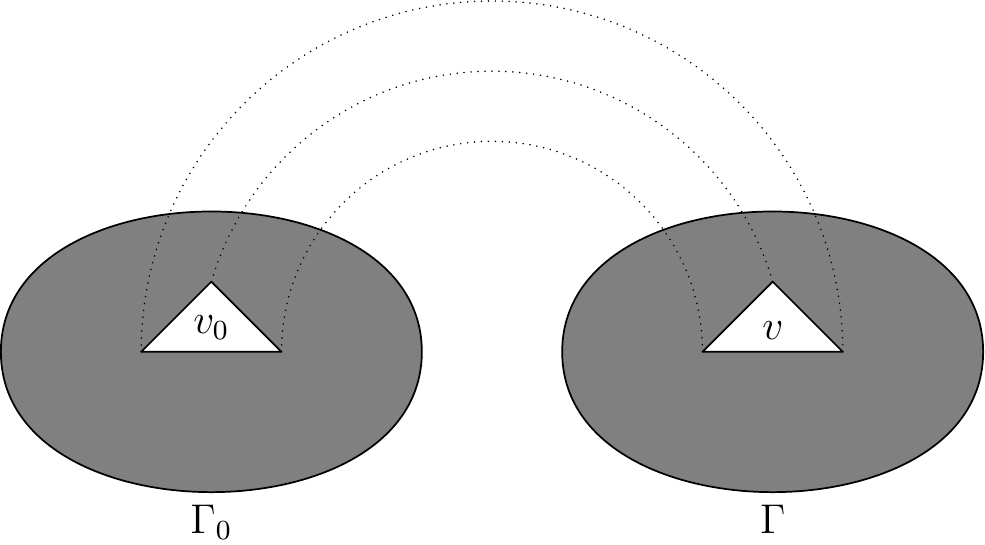}}\right\rangle=0.
\end{equation}
From an algebraic point of view, Schwinger-Dyson equations are constraints on the generating function ${\cal W}[\lambda_{\Gamma}]$ which can be written as 
\begin{equation}
{\cal L}_{(\Gamma_{0},\overline{v}_{0})}{\cal W}[\lambda_{\Gamma}]=0.
\end{equation}
The differential operators form a Lie algebra representing the Lie algebra of infinitesimal change of variables. For $D=2$, tensor models reduce to non hermitian matrix models, graphs $\Gamma$ are necklaces and the operators ${\cal L}_{n}$ define the WItt algebra $[{\cal L}_{m},{\cal L}_{n}]={\cal L}_{m-n}$. In the $D>2$, this algebra can be associated with a Hopf algebra of graphs of the Connes-Kreimer type.
 
\section{Lie and Hopf algebras of graphs} 
 
 A Hopf algebra ${\cal H}$ is an associative algebra together with a counit $\epsilon: {\cal H}\rightarrow {\Bbb R}$, a coproduct $\Delta: {\cal H}\rightarrow {\cal H}\otimes{\cal H}$ and an antipode $S: {\cal H}\rightarrow {\cal H}$ obeying some axioms (see for instance the book\cite{Chari}). Let us define ${\cal H}_{D}$ as the commutative algebra generated by  pairs $(\Gamma,v)$ where $\Gamma$ is a connected bipartite graph $\Gamma$ with vertices of order $D$ and a proper edge coloring with $D$ colors and $v\in\Gamma$ a vertex. The coproduct is of the Connes-Kreimer type\cite{CK} 
\begin{equation}
\Delta(\Gamma,v)=(\Gamma,v)\otimes 1+1\otimes(\Gamma,v)+\hskip-1cm\sum_{\Gamma_{n}\mathrm{\,disjoint\, subgraphs},\,v\notin\Gamma_{n}\atop \mathrm{\,with\,}D\,\mathrm{legs\,with\,different\,colors}}\hskip-0.4cm
\bigg(\prod_{n}(\widehat{\,\Gamma}_{n},v_{n})\bigg)\otimes\bigg( \Gamma\!\!\bigg/\!\!\prod_{n}\Gamma_{n}\,,v\bigg)
\end{equation} 
\noindent
$\Gamma_{n}$ is assumed to be connected and necessarily has all external legs attached to either black or white vertices. We define $\widehat{\Gamma}_{n}$ by attaching an extra vertex $v_{n}$ of the opposite color to the external legs of $\Gamma_{n}$. The reduced graph  $ \Gamma\big/\prod_{n}\Gamma_{n}$ is obtained by shrinking each $\Gamma_{n}$ to a single vertex in $\Gamma$.
 
 For any commutative Hopf algebra, the set of characters (i.e. linear and multiplicative maps from ${\cal H}$ to ${\Bbb C}$) form a group for the convolution product $\alpha\ast\beta=(\alpha\otimes\beta)\circ\Delta$ with unit $\epsilon$ and inverse $\alpha^{-1}=\alpha\circ S$.  Thus, characters of ${\cal H}_{D}$  define a group whose elements are fully described by their values on the generators $(\Gamma,v)$. The Lie algebra of this group is the Lie algebra of constraints of tensor models. Its generators are indexed by pairs $(\Gamma,v)$ and its bracket is the linearized form of the bracket derived from the convolution product.

\section{Effective action in colored tensor models}

 In the context of tensor models, the analogue of the Wilsonian effective action is defined as 
\begin{align}
S_{t}[T,\overline{T}]&=-\log \int \underbrace{[D\overline{Q}DQ]_{K_{t}}}_{\mbox{\tiny Gau\ss ian measure}}\exp-S_{0}[T+Q,\overline{T}+\overline{Q}]
\end{align}
with $T$ a fixed tensor and $Q$ a fluctuating one.  The latter is integrated out with a Gau\ss ian weight $K_{t}$ that depends on a parameter $t$ analogous to the floating infrared cut-off in quantum field theory. In the combinatorial model studied here, we focus on counting graphs instead of deriving scaling laws. Therefore, we choose the simplest Gau\ss ian weight $K_{t}(M,\overline{M})=t^{-1}\mbox{Tr}_{\mathrm{dipole}}\big[T,\overline{T}\big]$. The effective action is expanded over all colored bipartite graphs (included disconnected ones) with vertices of valence $D$
\begin{equation*}
S_{t}[T,\overline{T}]=\sum_{\Gamma}\frac{\lambda_{\Gamma}(t)}{C_{\Gamma}}\mbox{Tr}_{\Gamma}(T,\overline{T})\
\end{equation*}
For example, in the case of $2d$ gravity and matrix models, the expansion reads
\begin{align}
 S[M,M^{\dagger}]=\sum_{\left\{n_{k}\right\}}\lambda_{\left\{n_{k}\right\}}\prod_{k}\frac{1}{k^{n_{k}}n_{k}!}\big[\mbox{Tr}(MM^{\dagger})^{k}\big]^{n_{k}}
\end{align}
The effective action obeys a differential equation analogous to the Wilsonian flow equation in quantum field theory
 \begin{align}
 \frac{d\lambda_{\Gamma}}{dt}&=\sum_{0\leq k\leq D}\sum_{\mbox{\tiny edge cuts}\atop \left\{e_{1},\dots,e_{k}\right\} }\bigg\{ N^{D-k}\lambda_{\Gamma-\left\{e_{1},\dots,e_{k}\right\}}
 +\sum_{\Gamma_{1}\cup\Gamma_{2}=\Gamma-\left\{e_{1},\dots,e_{k}\right\} }\lambda_{\Gamma_{1}}\lambda_{\Gamma_{2}}\bigg\}
 \end{align}
The sum runs over all edge cuts of $k$ edges with different colors. The graph $\Gamma-\left\{e_{1},\dots,e_{k}\right\}$ is obtained by  cutting $k$ edges with different colors and attaching two new vertices to the resulting free half-edges. For instance,

\medskip  
 
\hskip-0.8cm
{\begin{tabular}{ccc} 
{\parbox{1.8cm}{\includegraphics[width=1.8cm]{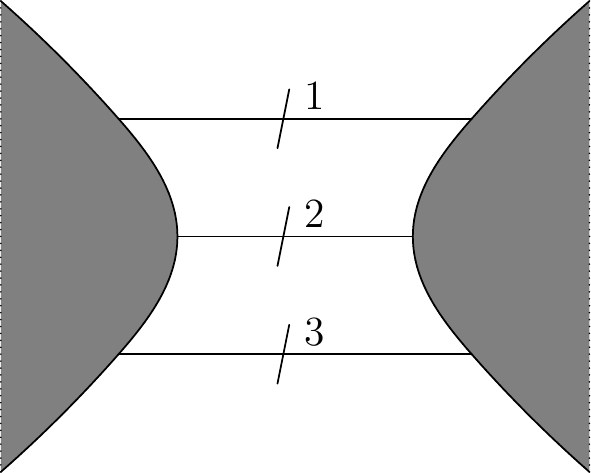}}}{$\,\rightarrow\,$\parbox{3cm}{\includegraphics[width=3cm]{cut1.pdf}}}&&
{\parbox{1.8cm}{\includegraphics[width=1.8cm]{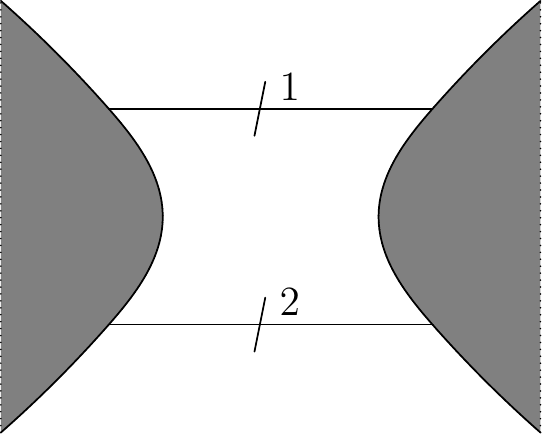}}}{$\,\rightarrow\,$\parbox{3cm}{\includegraphics[width=3cm]{cut3.pdf}}}\\
{cut with $D=3$ colors}&&{ cut with $D-1=2$ colors}\\
\end{tabular} }

\medskip

Note that these operations play a important role in the proof of the renormalizabilty of some colored models\cite{renormalizability}.

From a geometrical perspective, this equation can be depicted as

 {\centerline{\includegraphics[width=9cm]{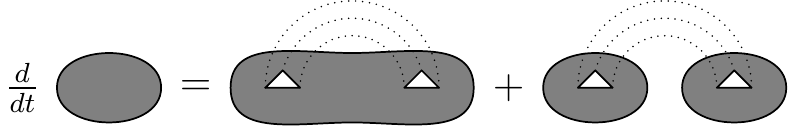}.}}
 
 In this equation, the effective action is a sum over connected Feynman graph of the tensor model. Therefore, it may be understood as a sum over connected $(D\!+\!1)$-dimensional triangulations, whose boundary may be disconnected. Then, the evolution of such a triangulation results either from an identification of two $D$-simplexes in the boundary of the same connected $(D\!+\!1)$-dimensional triangulation or in different $(D\!+\!1)$-dimensional triangulations.

\section{Extension to group field theory}

Group field theories\cite{Freidel} are particular quantum field theories defined on several copies of a group $G$ which may be $\mathrm{SU}(2)$, $\mathrm{SU}(2)\times \mathrm{SU}(2)$ or $\mathrm{SL}(2,{\Bbb C})$. In the sequel, we restrict our attention to $\mathrm{SU}(2)$. Their perturbative yield a sum over  $(D\!+\!1)$-dimensional triangulations weighted by the corresponding spin foam amplitudes appearing in quantum gravity\cite{Rovelli}. The graphs $\Gamma$ are at the basis of the loop quantum gravity boundary Hilbert spaces  ${\cal H}_{\Gamma}=\mathrm{L}^{2}(\mathrm{SU(2)}^{E/V})$ of gauge invariant functions $\Psi(h_{e})=\Psi(g^{}_{s(e)}h_{e}g_{t(e)}^{-1})$  with $s(e)$ and $t(e)$ source and target of the edge $e$.
The group field theory generating function analogous to \eqref{generating} reads
\begin{multline}
{\cal W}[\Psi_{\Gamma}]=\log\int[D\Phi]
\exp\Big\{\sum_{\Gamma}\int \!\!dh_{e}dg_{v,e}\\\, \frac{\Psi_{\Gamma}(h_{e})}{C_{\Gamma}}
\prod_{e}\delta\big(h^{}_{e}\,g_{s(e),e}^{-1}g^{}_{t(e),e}\big)\prod_{v}\Phi(g_{v,e_{1}},\cdots,g_{v,e_{D}})\Big\}
\end{multline}}
 A convenient basis of ${\cal H}_{\Gamma}$ is provided by spin networks $s=(\Gamma, j_{e},i_{v})$ constructed by associating a spin to every edge and to every vertex an intertwiner between all the spin associated to the edges incident to that vertex. Therefore, all the previous formalism can be extended mutatis mutandis to group field theory replacing graphs by spin networks.

\section*{Acknowledgements}

The author warmly thanks Vincent Rivasseau for his invitation to this conference and Yongge Ma for his hospitality at Beijing Normal University.

\end{document}